\documentclass[twocolumn,prd,groupedaddress,nofootinbib,showpacs]
{revtex4}

\usepackage{latexsym}
\usepackage{amssymb}
\usepackage{graphicx}
\usepackage{dcolumn}
\usepackage{bm}

\begin{document}

\title{The Hamiltonian BRST quantization of a noncommutative  nonabelian gauge theory
and its Seiberg-Witten map}
\author{Ricardo Amorim$^a$ and Franz A. Farias$^b$}

\affiliation{Instituto de F\'{\i}sica\\
Universidade Federal do Rio de Janeiro\\
Caixa Postal 68528, RJ 21941-972 -- Brazil}

\date{\today}

\begin{abstract}
We consider the Hamiltonian BRST   quantization of a
noncommutative non abelian  gauge theory. 
The Seiberg-Witten map of all  phase-space variables, including multipliers, 
ghosts and their momenta, is  given in first order in the noncommutative parameter $\theta$.
We show that there exists a complete consistence between the gauge structures of the original
and of the mapped theories, 
derived in a canonical way, once we appropriately choose the map solutions.
\end{abstract}

\maketitle

\section{Introduction}

\renewcommand{\theequation}{1.\arabic{equation}}
\setcounter{equation}{0}

Since the pioneer paper \cite{Snyder} about the
noncommutative structure for spacetime coordinates as an attempt
to introduce a natural ultraviolet cutoff for relativistic field
theories, a great  amount of work has been done concerning
noncommutative geometry  and noncommutative field
theories \cite{REVIEW}. In recent years, strong motivations for
further developing these subjects have appeared in the context of
string theories. It was shown that noncommutativity then naturally
arises in the effective action of open strings in the presence of
magnetic fields \cite{SW}. Due to nocommutatitity, the spacetime
coordinates  $x^\mu$ are  replaced by the Hermitian generators
$\hat x^\mu$ of a noncommutative $C^*$-algebra over spacetime
functions satisfying

\begin{equation}
[\hat x^\mu,\hat x^\nu]=i\theta^{\mu\nu}
\end{equation}

\noindent where $\theta^{\mu\nu}$, in the simplest description, is
a constant antisymmetric $D\times D$ matrix,  $D$ being the
spacetime dimension. Constructing quantum field theories starting
from these ingredients is a  difficult program. However, it is
possible to use the Weyl-Moyal ideas that relate operators to
classical functions with the use of the so called Weyl
transformations \cite{Weyl}. Instead of working with noncommuting
functions of the operators $\hat x^\mu$, it is then possible to
perform the appropriate calculations by using usual functions of
$x^\mu$. The price to be paid is the deformation of the  usual
commutative product to the noncommutative Moyal star product

\begin{equation}
\phi _{1}(x)\star \phi _{2}(x)=\exp \,\left( \frac{i}{2}
\theta ^{\mu \nu}\partial _{\mu }^{x}\partial _{\nu }^{y}\right) \,
\phi _{1}(x)\phi_{2}(y)|_{x=y}
\label{01}
\end{equation}

\noindent As can be verified, the space-time integral of the Moyal
product of two fields is the same as the usual one, provided we
discard boundary terms. So the noncommutativity affects  the
vertices in the action. These possibilities imply in several interesting features of
noncommutative quantum field theories \cite{REVIEW,SW}. 
\medskip

It is not difficult to deform  gauge
theories in order to get actions which are invariant under gauge
transformations associated with the Moyal structure. The form of
the gauge transformations imply, however, that the algebra must
close not only under commutation but also under anticommutation.
This usually makes $U(N)$ to be chosen as the symmetry group of
noncommutative Yang-Mills theories in place of $SU(N)$, although other symmetry groups can also be considered
\cite{Bonora}. It is possible, also,
to let the connections  take values in the enveloping algebra
of an arbitrary symmetry group \cite{Wess}. Once we chose a representation for such a group,
given for instance by $n\times n$ matrices, the corresponding enveloping algebra can 
be properly spanned by the
$u(n)$ generators in the representation  also given by  $n\times n$ hermitian matrices, since they form a basis
for that vector space. In this way we can 
consider   arbitrary symmetry groups, in a given representation,  with enveloping algebra spanned by  $u(n)$. 
\bigskip

The classical Lagrangian treatment of noncommutative Yang-Mills
theories poses no formal problems regarding  the specific values
the components of $\theta^{\mu\nu}$ can take. However, even a
classical Hamiltonian treatment depends strongly if the
noncommutative parameter $\theta ^{0i}$ vanishes or not
\cite{Hamiltonian,Banerjee,AF} . In the last case we
have an arbitrarily higher order derivative theory which has to be
treated with non canonical means. At quantum level this same
condition breaks unitarity. 
Due to the fact that $\theta^{\mu\nu}$ is a constant matrix, 
Lorentz invariance is lost in any case.
\bigskip

In this work, after reviewing  the Hamiltonian treatment  of the
gauge sector of a
noncommutative gauge theory where the connections take
values in a $u(N)$ algebra, as previously presented in \cite{AF}, and
in the   unitary case where
$\theta ^{0i}=0$,
we construct the Seiberg-Witten map for the phase space variables. 
We show that the specific solution
for the map of the momentum found in \cite{Banerjee} is actually the one
that permits us to consistently generate the constraints that 
give the appropriate gauge structure for the mapped theory. In this way 
we prove that the noncommutative $u(N)$ symmetry algebra can be actually generated from an underlying theory
which presents a commutative $u(N)$ algebra, both structures generated 
canonically via gauge generators acting with the aid of a Poisson bracket 
structure.
After that the BRST \cite{BRST} treatment of the
noncommutative $u(N)$ gauge theory is considered. In this context, we prove that there are no
structure functions of higher orders, which permits to construct
in a simple way the  appropriate extended phase space containing
ghosts, their momenta, trivial pairs and all of the BFV-BRST machinery
to generate nilpotent BRST
transformations, a Hamiltonian path integral with convenient
measure and appropriate gauge fixing. 
At this level, we consider 
again the Seiberg-Witten map including now the variables
of the extended phase, showing  its consistency also
at the level of BRST transformations. Some  results previously derived with the use of cohomological
techniques \cite{SW-BRST} are reproduced, but new results are here presented  as the
mapping of  ghost momenta, gauge generators ,  BRST charges  and extended actions, 
as far we know, by the first time.
\medskip

This work is organized as follows:
In Section II we present a brief review of some aspects already
treated in \cite{AF} in order to establish notation and
conventions. Section III is devoted to construct the Hamiltonian Seiberg-Witten map of the 
noncommutative $u(N)$ theory, and the canonical gauge structure of the mapped theory is displayed, 
showing that  it is canonically consistent
with the one of the original theory. In Section IV the BRST formalism
is applied to the original noncommutative theory, discussing its gauge structure,
BRST transformations and the path integral quantization, with appropriate gauge fixing. We  consider the 
Seiberg-Witten map of that
extended theory in Section V.
At last, in Section VI, we present some final remarks. In  an appendix we collect some results useful for  
calculations performed through the work.

\section{Hamiltonian description}

\renewcommand{\theequation}{2.\arabic{equation}}
\setcounter{equation}{0}

Accordingly to what has been discussed in the previous section, we 
start, without lost of generality, from the ( Lagrangian ) action which describes the gauge 
sector of a
noncommutative Yang-Mills theory, which can be written as

\begin{equation}
S=-\,\frac{1}{2}\,tr\int d^{4}x\,F_{\mu \nu }\star F^{\mu \nu }  \label{2.1}
\end{equation}

\noindent Here the curvature tensor is defined by

\begin{eqnarray}
F_{\mu \nu } &=&\partial _{\mu }A_{\nu }-
\partial _{\nu }A_{\mu }-i\,(A_{\mu
}\star A_{\nu }-A_{\nu }\star A_{\mu })  \nonumber \\
&\equiv &\partial _{\mu }A_{\nu }-\partial _{\nu }A_{\mu
}-i\,[A_{\mu } \buildrel\star\over, A_{\nu }]  \label{2.2}
\end{eqnarray}

\noindent and the connections take values in the $u(N)$ algebra, with
generators $T^{A}$, assumed to be normalized as

\begin{equation}
tr(T^{A}T^{B})={\frac{1}{2}}\delta ^{AB}  \label{2.3}
\end{equation}

They not only form a Lie algebra but also close under anticommutation:

\begin{eqnarray}
\lbrack T^{A},T^{B}] &=&if^{ABC}T^{C}  \nonumber \\
\{T^{A},T^{B}\} &=&d^{ABC}T^{C}  \label{2.4}
\end{eqnarray}

In the above expressions, we take $f^{ABC}$ and $d^{ABC}$ as
completely antisymmetric and completely symmetric respectively. From the
above equations, it is possible to derive the momenta conjugate to 
$A_{\mu}^{B}$ as

\begin{equation}
\Pi _{\mu }^{B}=\frac{\partial L}{{\partial }\dot{A}^{\mu B}}=F_{\mu 0}^{B}
\label{2.5}
\end{equation}

\noindent There are primary constraints

\begin{equation}
T_{1}^{A}=\Pi _{0}^{A}  \label{2.6}
\end{equation}

\noindent and the primary Hamiltonian is given by

\begin{eqnarray}
H_p&=&\int d^3x\,( {\frac{1}{2}}\Pi^{iB} \Pi^{iB} \nonumber \\
&+&{\frac{1}{4}}F_{ij}^B\, F^{ijB}-(D_i\Pi^i)^B A^{0B}+
\Lambda^{1B}T_1^B)  \label{2.7}
\end{eqnarray}

\noindent where
\begin{eqnarray}
(D_{i}\Pi ^{i})^{B}&=&\left(\partial_i\Pi^i-i[A_i\buildrel\star\over,\Pi^i]\right)^B\nonumber\\
&=&
\partial _{i}\Pi ^{iB}+{\frac{1}{2}}f^{BCD}\{A_{i}^{C}\buildrel\star\over,
 \Pi ^{iD}\}
\nonumber\\
&-&{\frac{i}{2}}d^{BCD}[A_{i}^{C}\buildrel\star\over, \Pi ^{iD}]
\label{2.8}
\end{eqnarray}
\medskip

With the aid of the Poisson brackets

\begin{eqnarray}
& &\{X(x),Y(y)\}_{PB}\nonumber\\
& &=\int d^{3}z\,\left( \frac{\delta X(x)}{\delta A_{\mu
}^{C}(z)}\frac{\delta Y(y)}{\delta \Pi ^{\mu C}(z)}-\frac{\delta Y(y)}{%
\delta A_{\mu }^{C}(z)}\frac{\delta X(x)}{\delta \pi ^{\mu C}(z)}\right)\nonumber\\
\label{2.9}
\end{eqnarray}

\noindent with $x^{0}=y^{0}=z^{0}$, the time evolution of the
primary constraints  imply, as usual, the secondary constraint

\begin{equation}
\{T_{1},H_{p}\}_{PB}=(D_{i}\Pi ^{i})\equiv T_{2}  \label{2.10}
\end{equation}

\noindent It is easy to verify that the constraint $T_1^A$ satisfies the
abelian algebra

\begin{eqnarray}
\{T_{1}^{A}(x),T_{1}^{B}(y)\}_{PB} &=&0  \nonumber \\
\{T_{1}^{A}(x),T_{2}^{B}(y)\}_{PB} &=&0  \label{2.11}
\end{eqnarray}

\noindent but only after a bit longer calculation it is possible to show
that $T_{2}^{A}$ closes in an algebra with itself \cite{AF}:

\begin{eqnarray}
\{T_{2}^{A}(x)&,&T_{2}^{B}(y)\}_{PB}={\frac{1}{2}}f^{ABC}\{\delta (x-y)\buildrel\star\over,
T_{2}^{C}(x)\}\nonumber\\
& &-{\frac{i}{2}}d^{ABC}[\delta (x-y)\buildrel\star\over, T_{2}^{C}(x)]
\label{2.12}
\end{eqnarray}

One can also prove  that

\begin{eqnarray}
\{T_{2}^{A},H\}_{PB} &=&{\frac{1}{2}}f^{ABC}\{\Lambda ^{2B}-A^{0B}\buildrel\star\over,
T_{2}^{C}\}
\nonumber\\
&-&{\frac{i}{2}}d^{ABC}[\Lambda ^{2B}-A^{0B}\buildrel\star\over, T_{2}^{C}]
\nonumber \\
&=&-i[\Lambda ^{2}-A^{0}\buildrel\star\over, T_{2}]^{A}  \label{2.13}
\end{eqnarray}

\noindent and consequently no more constraints are produced. In the above
equation,

\begin{equation}  \label{2.14}
H=H_p+2\,Tr\int d^3x\Lambda^{2}T_2
\end{equation}

\noindent is the first class Hamiltonian.
\bigskip

Now, it is not difficult to show that the gauge invariance of the first order action

\begin{eqnarray}
S_{FO}&=&\int d^4x \,\Pi^{\mu B}\,\dot A_\mu^B - \int dx^0 H  \nonumber \\
&=&tr\int d^4x \, (2\Pi^{\mu}\,\dot A_\mu-\Pi^i\Pi^i-{\frac{1}{2}}%
F_{ij}F^{ij}
\nonumber\\
&-&2T_2(\Lambda^2-A^{0})-2T_1\,\Lambda^1)  \label{2.15}
\end{eqnarray}

\noindent can be achieved with the aid of the gauge generator

\begin{equation}
G=-2\,tr\int d^{3}x\,(\epsilon ^{1}T_{1}+\epsilon ^{2}T_{2})
\label{2.16}
\end{equation}

\noindent which acts canonically on the phase space variables $Y$ through
$\delta Y=\{Y,G\}_{PB}$ to produce the gauge transformations

\begin{eqnarray}
\delta A^{0} &=&\epsilon ^{1}  \nonumber \\
\delta A_{i} &=&D_{i}\epsilon ^{2}  \nonumber \\
\delta \Pi ^{0} &=&0  \nonumber \\
\delta \Pi _{i} &=&i[\epsilon ^{2}\buildrel\star\over,\Pi _{i}]  
\label{2.17}
\end{eqnarray}

Indeed (\ref{2.15}) is invariant under (\ref{2.17}) if we also assume that

\begin{eqnarray}
\delta \Lambda ^{1} &=&\dot{\epsilon}^{1}  \nonumber \\
\delta \Lambda ^{2} &=&\epsilon ^{1}-\dot{\epsilon}^{2}+
i[\Lambda^{2}-A^{0}\buildrel\star\over, \epsilon ^{2}]
\label{2.18}
\end{eqnarray}

As expected, the redefinition

\begin{equation}
A^0 \rightarrow \tilde A^0=A^0-\Lambda^2
\label{2.18b}
\end{equation}

\noindent permits to write the gauge transformations involving
$\tilde A_0$ and $A_i$ in the covariant way, the gauge
transformation of the connections defined as $\delta
A_\mu=D_\mu\epsilon^2$. It is useful to note that the
transformation of $\Pi_\mu$ is consistent with the identification
(\ref{2.5}), since from (\ref{2.2}) and (\ref{2.17}) we arrive
directly to

\begin{equation}
\delta F_{\mu\nu}=i[\epsilon^2 \buildrel\star\over, F_{\mu\nu}]
\label{2.19}
\end{equation}

\noindent when one uses (\ref{2.18b}).

\section{ The Seiberg-Witten map}

\renewcommand{\theequation}{3.\arabic{equation}}
\setcounter{equation}{0}

The gauge transformations appearing in (\ref{2.17},\,\ref{2.18}), here generically written as 
$\delta Y$, close in
the algebra

\begin{equation}
[\delta_1,\delta_2]\,Y=\delta_3\, Y
\label{SW1}
\end{equation}

\noindent where $Y$ represents any one of the fields appearing in  those equations.
As can be verified, the composition rule for the parameters is given by

\begin{eqnarray}
\epsilon^1_3&=&0\nonumber\\
\epsilon^2_3&=&i[\epsilon^2_1\buildrel\star\over,\epsilon^2_2]
\label{SW2}
\end{eqnarray}

\noindent which means that the gauge sector involving $\epsilon^1$
is abelian and the noncommutativity is actually associated with
the gauge sector involving $\epsilon^2$. 
This means, for instance, that $[\delta_1,\delta_2]\,A^0=0$, or
$[\delta_1,\delta_2]\,A_i=D_i\epsilon^2_3$, with $\epsilon^2_3$ given by (\ref{SW2}).

\medskip
Now, for a possibly underlying commutative
 gauge theory, where the corresponding phase space variables 
written here with small letters, we would have gauge transformations  $\bar\delta y$ and
algebras similar to those listed above, but replacing the Moyal
commutators by usual ones. Specifically,

\begin{equation}
[\bar\delta_1,\bar\delta_2]\,y=\bar\delta_3\, y
\label{SW3}
\end{equation}

\noindent with $y$  representing the commutative fields and the
corresponding gauge parameters  designated by $\alpha$ in place of $\epsilon$. They must
obey the composition rule

\begin{eqnarray}
\alpha^1_3&=&0\nonumber\\
\alpha^2_3&=&i[\alpha^2_1,\alpha^2_2]
\label{SW4}
\end{eqnarray}

\noindent since it naturally follows from the 
commutative limit of (\ref{SW2}).

The basic idea in the Seiberg-Witten map is to write the
noncommutative fields $Y$ as functions of the 
commutative fields $y$. It is  assumed as well in the noncommutative
parameters $\epsilon$ a dependence on the commutative fields $y$
and parameters $\alpha$ in such a way that
$\bar\delta Y[y]=\delta Y$. The form of the dependence of $\epsilon$ 
 on the commutative fields $y$
and parameters $\alpha$ is determined when one also assumes that

\begin{equation}
[\bar\delta_1,\bar\delta_2]\,Y[y]=\bar\delta_3\, Y[y]
\label{SW5}
\end{equation}

\noindent using the composition rule given by (\ref{SW4}).
With these considerations  and taking in account
(\ref{2.18b}), we can write (\ref{SW5}) in detail as

\begin{widetext}

\begin{eqnarray}
& &\left[\,\bar\delta_1,\bar\delta_2\,\right]\, \Lambda^1[y] =0
\nonumber \\
& &\left[\,\bar\delta_1,\bar\delta_2\,\right]\,  \Pi_0[y] =0
\nonumber\\
& &\left[\,\bar\delta_1,\bar\delta_2\,\right]\,  \, \Pi_i[y]\,=
\,i\left[\bar\delta_1\epsilon^2_2[y]-
\bar\delta_2\epsilon^2_1[y]+
i\left[\epsilon^2_2[y]\buildrel\star\over,\epsilon^2_1[y]\right]
\buildrel\star\over,\Pi_i[y]\right]
\nonumber\\
& &\phantom{\left[\,\bar\delta_1,\bar\delta_2\,\right]\, \Pi_i[y]\,}=
i\left[\epsilon^2_3[y]\buildrel\star\over,\Pi_i[y]\right]\nonumber\\
& &\left[\,\bar\delta_1,\bar\delta_2\,\right]\,  A_\mu [y]=D_\mu\left(
\,\bar\delta_1\epsilon^2_2[y]-
\bar\delta_2\epsilon^2_1[y]+
i\left[\epsilon^2_2[y]\buildrel\star\over,\epsilon^2_1[y]\right]\,\right)
\nonumber\\
& &\phantom{\left[\,\bar\delta_1,\bar\delta_2\,\right]\, \Pi_i[y]\,}=
D_\mu\epsilon^2_3[y]
\nonumber\\
\label{SW6}
\end{eqnarray}

\end{widetext}

\noindent where $\epsilon^2_3[y]$ is a shorthand notation for $\epsilon^2[\alpha^2_3,y]$.
\medskip

Let us consider with some detail the gauge sector which is non
trivial. To simplify the notation, let us  suppress the
superior index $2$ denoting the class of $\alpha$ or $\epsilon$
parameters in the forthcoming equations. We see that the last two
equations in (\ref{SW6}), involving the transformation of $\Pi_i$ and $A_i$,
imply that

\begin{equation}
\label{SW7}
\epsilon_3[y]=
\bar\delta_1\epsilon_2[y]-
\bar\delta_2\epsilon_1[y]+
i\left[\epsilon_2[y]\buildrel\star\over,\epsilon_1[y]\right]
\end{equation}

\noindent in place of (\ref{SW2}). 
Now, (\ref{SW6}) is exactly the equation appearing for instance in
\cite{Wess} whose solution, at first order in $\theta$, is given
by

\begin{eqnarray}
\epsilon_\alpha[y]&=&\alpha+  \epsilon^{(1)} +O(\theta^2)\nonumber\\
&=&\alpha+\frac{1}{4}\theta^{ij}\left\{\partial_i\alpha,a_j\right\}+O(\theta^2)
\label{SW8}
\end{eqnarray}

\noindent which defines $\epsilon^{(1)}$, and so it is enough to consider a 
field dependence of the parameters in the noncommutative connections
$a_\mu $. 
The map of $A_\mu$, which also has been worked out in the literature,  
comes from  (\ref{2.17}-\ref{2.18b}), 
when one also considers (\ref{SW8}) and imposes that $\bar\delta A_\mu[y]=\delta A_\mu$. We get \cite{SW}

\begin{equation}
A_\mu[a]=a_\mu-\frac{1}{4}\theta^{kl}\left\{a_k,\partial_l a_\mu+f_{l\mu}\right\}+O(\theta^2)
\label{SW9}
\end{equation}

\noindent where

\begin{equation}
f_{\mu\nu}=\partial _{\mu }a_{\nu }-\partial _{\nu }a_{\mu}-i\,[a_{\mu } , a_{\nu }]
\label{SW10}
\end{equation}

\noindent is the usual commutative non abelian $u(N)$ Faraday tensor. Let us consider with a 
bit more of detail the map for $\Pi_i$. Defining

\begin{equation}
\Pi_i[y]=\pi_i+\Pi_i^{(1)}+0(\theta^2)      
\label{SW10b}
\end{equation}

\noindent 
we see from (\ref{2.17}) and the above equations, in $0(\theta^2)$, that

\begin{eqnarray}
\delta \Pi _{i} &=&i[\alpha+\epsilon^{(1)}\buildrel\star\over,\pi_i+\Pi_i^{(1)}]\nonumber\\
&=&i[\alpha,\pi_i]+i[\alpha,\Pi_i^{(1)}]
-\frac{1}{2}\theta^{kl}\{\partial_k\alpha,\partial_l\pi_i\}\nonumber\\
&+& \frac{i}{4}\theta^{kl}
\{[\partial_k\alpha,a_l],\pi_i\}
\label{SW10c}
\end{eqnarray}

\noindent and remembering that 
$\bar\delta \Pi_i[y]=\delta \Pi_i$ and that $\bar\delta \pi_i=i[\alpha,\pi_i]$, 
we arrive at an equation for $\Pi_i^{(1)}$, in $0(\theta ^2)$, given by

\begin{equation}
\bar \delta \Pi_i^{(1)}-i[\alpha,\Pi_i^{(1)}]= - \frac{1}{2}\theta^{kl}
\left(\{\partial_k\alpha,\partial_l\pi_i\}-\frac{i}{2}
\{[\partial_k\alpha,a_l],\pi_i\}\right)
\label{SW10d}
\end{equation}

There is only a finite number of terms that can be candidates for solving the
above equation, due to symmetry and dimensional arguments. First we note that
any linear combination of $\theta^{kl}\{f_{ik},\pi_l\}$, 
$\theta^{kl}\{f_{kl},\pi_i\}$ or $\theta_{ik}\{f^{kl},\pi_l\}$
will solve the homogeneous part of (\ref{SW10d}). After a long but directly calculation,
one can show that the inhomogeneous part 
of (\ref{SW10d}) can be precisely solved by the particular solution
$-\frac{1}{4}\theta^{kl}\{a_k,(\partial_l+D_l)\pi_i\}$. In this way we get the general solution

\begin{eqnarray}
\Pi_i[y]&=&\pi_i+c_1\theta^{kl}\{f_{ik},\pi_l\}\nonumber\\
&+&c_2\theta^{kl}\{f_{kl},\pi_i\}+c_3\theta_{ik}\{f^{kl},\pi_l\}\nonumber\\
&-&\frac{1}{4}\theta^{kl}\{a_k,(\partial_l+D_l)\pi_i\}
\label{SW10e}
\end{eqnarray}

\noindent for the Seiberg-Witten map of the momentum.
Now, it is useful to observe that
from (\ref{2.2},\ref{SW9}),  at first order in $\theta$ \cite{SW},

\begin{equation}
F_{\mu\nu}[y]=f_{\mu\nu}+\frac{1}{2}\theta^{kl}\{f_{\mu k},f_{\nu l}\}
-\frac{1}{4}\theta^{kl}\{a_k,(\partial_l+D_l)f_{\mu\nu}\}
\end{equation}

\noindent so there exists a map for $\Pi_i$ which is consistent with  (\ref{2.5}) and the above definition, 
 given by

\begin{eqnarray}
\Pi_i^\prime[y]&=&\pi_i-\frac{1}{2}\theta^{kl}\{f_{i k},\pi_l\}-
\frac{1}{4}\theta^{kl}\{a_k,(\partial_l+D_l)\pi_i\} \nonumber\\&+&0(\theta^2)
\label{SW12}
\end{eqnarray}

\noindent This is also in accordance with (\ref{SW10e}),
although this is not the only solution found in literature.  
\medskip

The map for $\Pi_i$ obtained in Ref. \cite{Banerjee}, for instance,
which is different from
(\ref{SW12}), is also a particular case of  (\ref{SW10e}). It  is given by

\begin{eqnarray}
\Pi_i[y]&=&\pi_i+\frac{1}{4}\theta^{kl}\{f_{kl},\pi_i\}+\frac{1}{2}\theta_{ik}\{f^{ kl},\pi_l\}\nonumber\\
&-&
\frac{1}{4}\theta^{kl}\{a_k,(\partial_l+D_l)\pi_i\} +0(\theta^2)
\label{SW12b}
\end{eqnarray}

\noindent Soon we will show that with this last choice we get a consistent canonical formulation,
when we also take in consideration de map of the Lagrange multipliers. 
\medskip

The map for
$\Lambda^1$ and $A^0$ is trivial, since the sector involving
$\epsilon^1$ is abelian. The same occurs with $\Pi^0$, since it is
invariant. We get

 \begin{eqnarray}
 A_0[y]&=&a_0\nonumber\\
 \Pi^0[y]&=&\pi^0\nonumber\\
 \Lambda^1[y]&=&\lambda^1
 \label{SW13}
 \end{eqnarray}

Naturally  (\ref{SW10e}) and the spatial part of (\ref{SW9})  are still valid   for the map of
$A_i[y]$  and $\Pi_i[y]$. 
Of course, from (\ref{SW9}) we get ( explicitly writing the tilde )

\begin{equation}
\tilde A_0[a]=\tilde a_0-\frac{1}{4}\theta^{kl}\left\{a_k,\partial_l \tilde a_0+\tilde f_{l0}\right\}+O(\theta^2)
\label{SW13b}
\end{equation}

\noindent By using (\ref{2.18b}), which implies not only  $\tilde A_0=A_0-\Lambda_2$
but also that $\tilde a_0=a_0 - \lambda_2$, and  remembering (\ref{SW13}),  we arrive at

\begin{equation}
\Lambda^2[y]=\lambda^2-\frac{1}{4}\theta^{kl}\left\{a_k,(\partial_l+D_l)\lambda^2-\partial_l a_0-\pi_l\right\}
+O(\theta^2)
\label{SW14}
\end{equation}

\noindent which completes  the Hamiltonian Seiberg-Witten map for all the pertinent variables.
The  corresponding Hamiltonian action can then be written in first order in $\theta$,
from  the full action (\ref{2.15}) and the map described above as

\begin{eqnarray}
S_{FO}[y]&=&tr\int d^4x \, (2\Pi^{\mu}[y]\,\dot A_\mu[y]-\Pi^i[y]\Pi^i[y]\nonumber\\
&-&{\frac{1}{2}}
F_{ij}[y]F^{ij}[y]
-2T_2[y](\Lambda^2[y]-A^{0}[y])\nonumber\\
&-&2T_1[y]\,\Lambda^1[y]) +0(\theta^2)\label{SW15}
\end{eqnarray}

\noindent It is, by construction,
invariant under the $\bar\delta$ variations, since the original Noether identities 
are not altered, by construction, under the Seiberg-Witten map.

\medskip

Now one should be able to show that
the natural constraints coming from the action described above not only are first class but generate the
comummutative $u(N)$ gauge transformations, designated by $\bar\delta$, 
for instance, in (\ref{SW3}). Actually, it is easy to show that

\begin{eqnarray}
\bar t_1&=& -\frac{\delta S_{FO}[y]}{\delta\lambda^1}\nonumber\\
&=&\pi^0\nonumber\\
\bar t_2&=& -\frac{\delta S_{FO}[y]}{\delta\lambda^2}\nonumber\\
&=&T_2[y] +\frac{1}{4}\theta^{kl}(\partial_l+D_l)\{t_2,a_k\}+O(\theta^2)
\label{SW16}
\end{eqnarray}

\noindent where  
$T_2[y]= \partial_i\Pi^i[y]- i[A_i[y]\,\buildrel\star\over, \Pi^i[y]]$ and 
$t_2=\partial_i\pi^i - i[a_i,\pi^i]$, as in (\ref{2.10}).
At this stage,  after a long calculation with the use of Bianchi and Jacobi identities 
and many cancelations,
we arrive at a simple relation involving both quantities above, which is given by

\begin{equation}
T_2[y]=t_2+\frac{1}{4}\theta^{kl}(\partial_k+D_k)\{a_l,t_2\}
\label{SW16b}
\end{equation}

\noindent when (\ref{SW12b}) is chosen as the Seiberg-Witten  map for the momentum.
This fact implies, via (\ref{SW16}), that
$\bar t_2=t_2$, which guarantees that the underlying gauge structure actually exists and 
is the one given by the commutative
$u(N)$ symmetry. Indeed, by defining the gauge generator

\begin{equation}
g=-2\,tr\int d^{3}x\,(\alpha ^{1}\bar t_{1}+\alpha ^{2} \bar t_{2}),
\label{SW17}
\end{equation}

\noindent it is   trivial to show, via $\bar\delta y=\{y,g\}_{PB}$, that

\begin{eqnarray}
\bar\delta a^{0} &=&\alpha ^{1}  \nonumber \\
\bar\delta a_{i} &=&D_{i}\alpha ^{2}  \nonumber \\
\bar\delta \pi ^{0} &=&0  \nonumber \\
\bar\delta \pi _{i} &=&i[\alpha ^{2}\,,\pi _{i}].  \label{SW18}
\end{eqnarray}

\noindent and a complete consistency between the canonical Hamiltonian formalism of 
the original noncommutative theory
and the one of the mapped commutative theory is achieved in $O(\theta^2)$. Although the results we have derived are
strictly valid in this order in $\theta$, we conjecture that the identity between $t_2$ and $\bar t_2$ probably
is valid
at all orders in $\theta$, 
since only with this identity we would guarantee that the underlying gauge  symmetry of the mapped theory 
presents the desired structure. This fact has actually been proved, in higher orders in $\theta$, 
in a Lagrangian
formalism context, exploring directly the form of the transformations \cite{REVIEW,SW,Wess}.

\bigskip

\section{ BRST quantization}
\renewcommand{\theequation}{4.\arabic{equation}}
\setcounter{equation}{0}

Once we have reviewed the classical aspects of the Hamiltonian
treatment of the noncommutative $u(N)$ gauge theory, we are ready
to consider its Hamiltonian BRST formulation \cite{Soroush}. This is the first step to derive the 
functional quantization of
the theory from a constructive point of view. Let us first
consider the full theory treated in section II. 
In the next section we will consider the BRST quantization of the
mapped theory obtained in the previous section. Accordingly to the
usual procedure adopted in the BFV-BRST quantization of usual
Yang-Mills (Y-M) theory \cite{BRST}, here we also  discard the $N^{2}$ pairs
$\left( A_{0},\Pi_{0}\right) $ absorbing $A_{0}$ in $\Lambda^2$ so
that $\left( A_i,\Pi_i\right) $ and the multipliers $\Lambda^2$
and their canonical momenta can be taken as the  dynamical variables of
the theory. The relevant algebraic structure to be considered is then
the one given by (\ref{2.12}). If we rewrite Eq. (\ref{2.12}) as

\begin{equation}
\left\{ T_{2}^{A}\left( x\right) ,T_{2}^{B}\left( y\right) \right\}_{PB}=
2\int d^{3}z\stackrel{(1)}{U^{ABC}}\left( x,y,z\right) T_{2}^{C}\left(
z\right)   \label{06a}
\end{equation}

\noindent the first order structure function $\stackrel{(1)}{U^{ABC}}$ is identified with

\begin{eqnarray}
\stackrel{(1)}{U^{ABC}}\left( x,y,z\right)  &=&\frac{1}{4}f^{ABC}\left\{
\delta \left( x-z\right) \buildrel\star\over, \delta \left( z-y\right) \right\}   \nonumber \\
&&+\frac{i}{4}d^{ABC}\left[ \delta \left( x-z\right) \buildrel\star\over, \delta \left(
z-y\right) \right] \nonumber\\
\label{07}
\end{eqnarray}

By  using the Jacobi identity, it can be proved that the existence
of non trivial second order structure functions depends on the
quantity

\begin{equation}
\stackrel{(1)}{D^{ABCD}}=\left\{ \stackrel{(1)}{U^{\text{ }\left[
AB\right\vert D\vert}},\stackrel{(0)}{U^{C]}}\right\} _{PB}+2\stackrel{(1)}{U^{%
\left[ AB\right\vert E\vert}}\stackrel{(1)}{U^{C]ED}}  \label{08}
\end{equation}

\noindent where $\stackrel{(0)}{U^{A}}\equiv T_{2}^{A}$ and
the integrations over intermediary variables are implicit.

Since $\stackrel{(1)}{U^{ABC}}$ does not depend on the phase space variables, the first term in
the right side of the above expression is trivially zero.
Therefore it  follows that

\begin{widetext}

\begin{eqnarray}
\stackrel{(1)}{D^{ABCD}}\left( x,y,z,w\right)  &=&2\stackrel{(1)}{U^{\left[
AB\right\vert E\vert}}\stackrel{(1)}{U^{C]ED}}
=\frac{1}{8}\int d^{3}u\left( f^{[AB|E\vert}f^{C]ED}\left\{ \delta \left(
x-z\right) \buildrel\star\over, \delta \left( z-u\right) \right\}
 \left\{ \delta \left(
u-y\right) \buildrel\star\over, \delta \left( u-w\right) \right\} \right.   \nonumber \\
&&+if^{[AB|E\vert}d^{C]ED}\left\{ \delta \left( x-z\right) \buildrel\star\over, \delta \left(
z-u\right) \right\} \left[ \delta \left( u-y\right) \buildrel\star\over, \delta \left(
u-w\right) \right]   \nonumber \\
&&+id^{[AB|E\vert}f^{C]ED}\left[ \delta \left( x-z\right) \buildrel\star\over, \delta \left(
z-u\right) \right] \left\{ \delta \left( u-y\right) \buildrel\star\over, \delta \left(
u-w\right) \right\}   \nonumber \\
&&-\left. d^{[AB|E\vert}d^{C]ED}\left[ \delta \left( x-z\right) \buildrel\star\over, \delta
\left( z-u\right) \right] \left[ \delta \left( u-y\right) \buildrel\star\over, \delta
\left( u-w\right) \right] \right) \nonumber\\
\label{09}
\end{eqnarray}

By using 

\begin{eqnarray}
f^{[AB|E|}f^{C]ED} &=&f^{DCE}f^{EAB}+f^{DAE}f^{EBC}+f^{DBE}f^{ECA}=0  \nonumber
\\
d^{[AB|E|}f^{C]ED} &=&f^{DCE}d^{EAB}+f^{DAE}d^{EBC}+f^{DBE}d^{ECA}=0
\label{12}
\end{eqnarray}

\noindent which are  consequence of Jacobi Identity,
 we  conclude that

\begin{equation}
\stackrel{(1)}{D^{ABCD}}\left( x,y,z,w\right) =0
\end{equation}

\noindent and  it is possible to choose the higher order structure
functions to vanish:

\begin{equation}
\stackrel{(n)}{U}=0\text{ \ \ for \ \ }n\geq 2  \label{14}
\end{equation}

At this point, we extend the original phase space by introducing the ghosts $C^A$ and their
momenta $\mathcal{P}^A$ in order to construct  the BRST operator as

\begin{eqnarray}
\Omega  &=&\int d^{3}xC ^{A}\left( x\right) T_{2}^{A}\left( x\right)
+\int d^{3}xd^{3}yd^{3}zC ^{B}\left( y\right) C ^{C}\left( x\right)
\times  \stackrel{(1)}{U^{ABC}}\left( x,y,z\right)
 \mathcal{P}^{A}\left( z\right)\nonumber\\
&=&2\,\mathrm{Tr}\int d^{3}x\left( C \left( x\right) T_{2}\left(
x\right) -i\left\{ C \left( x\right) \buildrel\star\over, C \left( x\right) \right\}
\mathcal{P}\left( x\right) \right)   \label{30}
\end{eqnarray}

\end{widetext}

\noindent 
To generate the BRST transformations and the dynamics in the extended phase
space, it is necessary to  extend the former
definition of the Poisson brackets in order to include
Grassmannian variables. As usual, we can write that

\begin{equation}
\{X,Y\} _{PB}  =
\frac{\partial ^{R}X}
{\partial Z^{\bar{A}}}\mathbf{C}^{\bar{A}\bar{B}}\frac{\partial ^{L}Y }
{\partial Z^{\bar{B}}}  \label{31a}
\end{equation}

\noindent where $Z^{\bar{A}}=\{A_i,\Pi_i,C,\mathcal{P}\}$ and intermediary integrations are
assumed. The symplectic matrix $\mathbf{C}^{\bar{A}\bar{B}}= \{z^{\bar{A}},z^{\bar{B}}\}_{PB}$ and the
 equal time Poisson Brackets
for the Grassmannian sector which do not vanish are given by
$\{C(x),\mathcal{P}(y)\}_{PB}=\{\mathcal{P}(x),C(y)\}_{PB}=-\delta(x-y)$.\bigskip

Now, a BRST transformation of an arbitrary quantity $X$ is generated via

\begin{equation}
s X =
\left\{ X,\Omega
\right\} _{PB}  \label{31}
\end{equation}

\noindent giving for the phase space variables, including ghosts,

\begin{eqnarray}
sA_{i} &=&-D_{i}C  \nonumber \\
s\Pi ^{i} &=&-i[C\buildrel\star\over,\Pi ^{i}]
\nonumber
\\
sC &=&\frac{i}{2} \{C\buildrel\star\over,C\}
\nonumber
\\
s\mathcal{P}
&=&-T_{2}+i\left\{C\buildrel\star\over,\mathcal{P}\right\} 
\label{32}
\end{eqnarray}

By noting that $s$ is an odd derivative acting from the right, it
is easy to demonstrate that actually it is nilpotent. For instance,

\begin{eqnarray}
s^{2}A_{i}&=&-s\left(\partial _{i}C-i\left[A_{i}\buildrel\star\over,C\right]\right)
\nonumber
\\
&=&-D_i(sC)+i\{D_iC\buildrel\star\over,C\}\nonumber
\\
&=&0
\nonumber
\end{eqnarray}

\noindent Also

\begin{eqnarray}
s^{2}\mathcal{P}&=&-sT_2+i[C \buildrel\star\over,s\mathcal{P}]-
i[sC \buildrel\star\over,\mathcal{P}]
\nonumber\\
&=&-sD_i\Pi^i
+i[C\buildrel\star\over,-D_i\Pi^i+i\{C\buildrel\star\over,
\mathcal{P}\} ]\nonumber\\
&+&
\frac{1}{2}[\{C\buildrel\star\over,C\}\buildrel\star\over,\mathcal{P}]
\nonumber\\
\end{eqnarray}

 \noindent which  vanishes identically when one inserts
 $sD_i\Pi^i=D_is\Pi^i-i[D_iC\buildrel\star\over,\Pi^i]$ in the above expression.
\bigskip

We follow the canonical approach which absorbs $A_0$ in $\Lambda_2$
(it is the reverse of  (\ref{2.18b})), discarding also the
variable $\Pi_0$, since the pair $A_0,\Pi_0$ is non-dynamical.  The momenta
conjugate to $\Lambda_2$ are also introduced and generate the constraints

\begin{equation}
\Gamma ^{A}=\frac{\partial \mathcal{L}}{\partial \dot{\Lambda_2}^{A}}%
\thickapprox 0  \label{73}
\end{equation}

\noindent
To simplify the notation, let us suppress in what follows the subscript $2$ in  $\Lambda_2 ^{A}$. 
With these considerations, we describe the phase space  with the variables
 $\{A_i,\Pi_i,\Lambda, \Gamma, C,\mathcal{P}\}$ 
and an additional pair of canonically conjugate ghosts $\Theta$ and $\bar C$ in order to implement the constraints in
the path integral.

The BRST operator for the non minimal space is redefined as

\begin{eqnarray}
\Omega &=&2\,\mathrm{Tr}\int d^{3}x( C\left( x\right) T_2\left(
x\right) \nonumber\\
&+&i\{ C\left( x\right) \buildrel\star\over, C\left( x\right)\} 
\mathcal{P}\left( x\right) -i\Theta \left( x\right) \Gamma \left( x\right)) 
\label{82}
\end{eqnarray}

\noindent and the extended action rewrites as below

\begin{equation}
S_{k}=\int d^{4}x\left( \dot{A}^{iA}\Pi _{i}^{A}+\dot{\Lambda}^{A}\Gamma
^{A}+\dot{C}^{A}\mathcal{P}^{A}+\dot{\Theta}^{A}\bar{C}^{A}-\mathcal{H}%
_{eff}\right)   \label{84}
\end{equation}

\noindent where $\mathcal{H}_{eff}=\mathcal{H}+\left\{ K,\Omega \right\} _{PB}$ and 
 the Hamiltonian density $\mathcal{H}$
being obtained from (\ref{2.7}) by letting $A_0$ and $\Lambda$ vanish. $K$ is the gauge fixing fermion.
\medskip

The BRST invariance of the action $S_{k}$ is demonstrated by the invariance
of the kinetic term since $\mathcal{H}+\left\{ K,\Omega \right\} _{PP}$ is an
extended invariant for $\mathcal{H}$ and so is BRST invariant. As can be verified, besides (\ref{32}) we get the BRST 
transformations of the two sets of trivial pairs

\begin{eqnarray}
s\Lambda  &=&-i\Theta \nonumber\\ 
s\Theta  &=&0  \nonumber\\ 
s\bar{C}&=&i\Gamma \nonumber\\ 
s\Gamma  &=&0  \label{55h}
\end{eqnarray}
\noindent which leads to  

\begin{equation}
s \,Tr\left( \dot{A}^{i}\Pi _{i}+\dot{\Lambda}\Gamma +\dot{C}\mathcal{P}+%
\dot{\Theta}\bar{C}\right) =-\partial _{i}\,Tr\left( \dot{C}\Pi^{i}\right) .
\end{equation}

\noindent which is a boundary term and can be discarded under the integral sign.

The gauge fixing procedure can be done as usually. Writing the gauge
fixing fermion as

\begin{equation}
K=2\,Tr(i\bar{C}\chi+\mathcal{P}\Lambda)  \label{95}
\end{equation}
where 

\begin{equation}
\chi=\partial ^{k}A_{k}  \label{96}
\end{equation}

\noindent we get

\begin{eqnarray}
\{K,\Omega\}_{PB}  &=&-2Tr(\Gamma\chi+i\bar C\partial_k D^k C\nonumber\\
&+&
(-T_2+i\{C \buildrel\star\over,\mathcal{P}\})\Lambda+i\mathcal{P}\Theta)
\label{99}
\end{eqnarray}

\noindent This permits to obtain an explicit form for $S_k$ according  to (\ref{84}), giving
\medskip

\begin{eqnarray}
S_{eff}&=&Tr\int d^{4}x( -\frac{1}{2}F_{\mu \nu }F^{\mu \nu}\nonumber\\
&+&2i\bar{C}\partial ^{\mu }D_{\mu }C +
2\Gamma\left( \partial ^{\mu}A_{\mu}\right))
\label{116}
\end{eqnarray}

\noindent after 
functionally integrating over $\Pi,\mathcal{P}$ and $\Theta$
and identifying $\Lambda$ and $A_0$. Other gauge choices are implemented in a similar way,
reproducing  the  results obtained if one starts directly from the Lagrangian formalism \cite{Soroush}.

\section{Mapping the extended theory}

\renewcommand{\theequation}{5.\arabic{equation}}
\setcounter{equation}{0}

As we have seen in Section III, 
the algebraic structure of the mapped theory is the one given by the $u(N)$ commutative theory. 
This implies that we can reproduce
every step developed in the previous section only by making trivial the structure due to the Moyal product, which
is obtained by letting $\theta$ vanish.
Following an obvious notation,
the extended phase space in now spanned by the commutative 
variables  $a_i,\pi_i,c,{p},\lambda,\gamma,\bar c$ and $\theta$, here generically denoted by $z^\Xi$. The BRST operator 
is obviously written as

\begin{eqnarray}
\omega&=&2\,\mathrm{Tr}\int d^{3}x( c \left( x\right) \bar t_{2}(x) \nonumber\\
&-&i\{ c( x), c ( x)\}
p( x)- i\theta\gamma)   \label{501}
\end{eqnarray}

\noindent which generates the BRST transformations

\begin{eqnarray}
\bar sa_{i} &=&-D_{i}c  \nonumber \\
\bar s\pi ^{i} &=&-i[c,\pi ^{i}]
\nonumber\\
\bar sc &=&\frac{i}{2} \{c,c\}
\nonumber\\
\bar sp &=&-t_{2}+i\left\{c,p\right\}\nonumber\\ 
\bar s\lambda  &=&-i\theta \nonumber\\ 
\bar s\theta  &=&0  \nonumber\\ 
\bar s\bar{c}&=&i\gamma \nonumber\\ 
\bar s\gamma  &=&0  \label{502}
\end{eqnarray}

Now we are in a position appropriate
to construct the Seiberg-Witten map in the extended phase space. By denoting by
$Z^\Xi$ the variables $A_\mu,\Pi_\mu,C,\mathcal{P},\Lambda,\Gamma,\bar C$ and $\Theta$,
we need to solve the relations

\begin{equation}
sZ=\bar sZ[z]
\label{503}
\end{equation}

\noindent in the same spirit of the one found in Section III. Consistence demands that the subset $Y$ of $Z$
must be mapped according the results already found in Section III. This implies that (see (\ref{SW8}) )

\begin{equation}
C[z]= c+\frac{1}{4}\theta^{ij}\left\{\partial_i c,a_j\right\}+O(\theta^2)
\label{504}
\end{equation}
\medskip

\noindent keeping the maps (\ref{SW9}) for $A_i[y]$, (\ref{SW12b}) for $\Pi_i[y]$ and (\ref{SW13}) for
$A_0[y]$ and $\Pi_0[y]$.
As can be verified, solution (\ref{504}) also consistently solve

\begin{equation}
s C=\bar s C[z]
\label{505}
\end{equation}

The next step is to found the map for $\mathcal{P}$. 
From (\ref{32}) and (\ref{SW16b})

\begin{eqnarray}
s\mathcal{P}&=&-T_{2}+i\{C\buildrel\star\over,\mathcal{P}\}\nonumber\\
&=&-t_2-\frac{1}{4}\theta^{kl}(\partial_k+D_k)\{a_l,t_2\}\nonumber\\
&+&
i\left\{ c+\frac{1}{4}\theta^{kl}\left\{\partial_k c,a_l\right\}  \buildrel\star\over,\mathcal{P}\right\}+
O(\theta^2)\label{506}
\end{eqnarray}

By writing  $\mathcal{P}=p+\mathcal{P}^{(1)}$ and remembering (\ref{502}) and (\ref{505}), we arrive at an 
equation for $\mathcal{P}^1$ given by

\begin{eqnarray}
\bar s\mathcal{P}^{(1)}[z]&-&i\{c,\mathcal{P}^{(1)}[z]\}\nonumber\\
&=&-\frac{1}{4}\theta^{kl}((\partial_k+D_k)\{a_l,t_2\}+2[\partial_k c,\partial_l p]\nonumber\\
&+&i\{\{\partial_k c,a_l\},p\})
+O(\theta^2)\label{507}
\end{eqnarray}

As can be verified after a long calculation, the solution of the above equation gives
a simple expression for $\mathcal{P}^{(1)}[z]$ and implies that

\begin{equation}
\mathcal{P}[z]=
p+\frac{1}{4}\theta^{kl}(\partial_k+D_k)\{a_l,p\}
\label{508}
\end{equation}

\noindent in $O(\theta^2)$. We observe, as expected, that all the obtained maps respect ghost and antighost degrees.
\medskip

The construction of the remaining maps for $\Lambda,\Gamma,\bar C$ and $\Theta$ is immediate, 
since they form trivial pairs. It is enough to identify these last 
quantities with the corresponding ones with small letters. Putting everything together, we see that we have succeeded in
solving
(\ref{503}) . This  implies that the 
extended action $S_k$ appearing in (\ref{84}) can be  mapped as well. Due to (\ref{503}) the mapped action
is BRST invariant, when the BRST transformations are given (\ref{502}).
Remaining points as
the gauge fixing fermion,  the measure  and the external sources to properly define the generating functional 
of the mapped theory will
not be considered here.

\section{Conclusion}

In this work we have  considered the Hamiltonian formalism concerning the gauge sector of a generic 
noncommutative gauge theory, whose enveloping algebra structure is embedded in a noncommutative
$u(N)$ algebra. We have succeeded in constructing, at first order in the noncommutative parameter 
$\theta$, its appropriate Hamiltonian Seiberg-Wittem map, showing the algebraic consistence between 
the gauge transformations of
both descriptions, the original and the mapped theory, generated canonically. To achieve this goal, 
it was necessary to choose the proper solutions of
the Seiberg-Witten map between the phase space variables.
We also have presented the BRST extensions of both theories,
generated via the action of BRST charges. 
The Seiberg-Witten map between those descriptions has been then constructed with the aid of the BRST transformations,
being in accordance with the results previously found.
The map has been consistently given for all the variables of the extended phase space, 
including trivial pairs, ghosts and their momenta, gauge generators, BRST charges, extended actions etc..
Some points regarding gauge fixing in the Hamiltonian path integral formalism of the original theory
have also been discussed.

\medskip

Acknowledgment: This work is supported in part by CAPES and CNPq (Brazilian
research agencies).

\appendix
\section{Some identities related to the Moyal product}

\renewcommand{\theequation}{A.\arabic{equation}}
\setcounter{equation}{0}

\begin{eqnarray*}
&&\int d^{4}x\phi _{1}\star \phi _{2} =\int d^{4}x\phi _{1}\phi _{2}=\int
d^{4}x\phi _{2}\star \phi _{1}  \label{B1}
  \nonumber \\
&&\left( \phi _{1}\star \phi _{2}\right) \star \phi _{3} =\phi _{1}\star
\left( \phi _{2}\star \phi _{3}\right) =\phi _{1}\star \phi _{2}\star \phi
_{3}  \label{B2} 
  \nonumber \\
&&\int d^{4}x\,\phi _{1}\star \phi _{2}\star \phi _{3} =\int d^{4}x\,\phi
_{2}\star \phi _{3}\star \phi _{1}=\int d^{4}x\,\phi _{3}\star \phi
_{1}\star \phi _{2}  \label{B3} 
  \nonumber \\
&&\phi (x)\star \delta (x-y) =\delta (x-y)\star \phi (y)  \label{B4} 
  \nonumber \\\nonumber \\
&&\phi (x)\star \partial _{\Lambda }^{x}\delta (x-y) =-\partial _{\Lambda
}^{y}\delta (x-y)\star \phi (y)  \label{B5} 
  \nonumber \\\nonumber \\
&&\lbrack \phi _{1},[\phi _{2},\phi _{3}]]+[\phi _{2},[\phi _{3},\phi
_{1}]]+[\phi _{3},[\phi _{1},\phi _{2}]] =0   
  \nonumber \\\nonumber \\
&&\lbrack \phi _{1}(x),[\phi _{2}(x),\delta (x-y)]] =[\phi _{2}(y),[\phi
_{1}(y),\delta (x-y)]]\nonumber\\  
\end{eqnarray*}

\end{document}